\documentclass[a4paper,fleqn,usenatbib]{mnras}

\usepackage{newtxtext,newtxmath}

\usepackage[T1]{fontenc}
\usepackage{ae,aecompl}

\usepackage{caption}

\usepackage{graphicx}	

\newcommand{\RNum}[1]{\uppercase\expandafter{\romannumeral #1\relax}}

\newcommand{\nh}{\hat{\mathbf{n}}}

\title[Point sources in full-sky interferometric maps]{An efficient method for removing point sources from full-sky radio interferometric maps}

\author[P. Berger et al.]{
Philippe Berger$^{1, 2}$\thanks{E-mail: pberger@cita.utoronto.ca},
Niels Oppermann$^{1, 3}$,
Ue-Li Pen$^{1, 4, 3, 2, 5}$,
and J. Richard Shaw$^{6}$
\\
$^{1}$Canadian Institute for Theoretical Astrophysics, 60 St. George St., Toronto, ON, M5S 3H8, Canada\\
$^{2}$Department of Physics, University of Toronto, 60 St George St, Toronto, ON, M5S 1A7, Canada \\
$^{3}$Dunlap Institute for Astronomy \& Astrophysics, University of Toronto, 50 St George St, Toronto, ON, M5S 3H4, Canada \\
$^{4}$Canadian Institute for Advanced Research, CIFAR Program in Gravitation and Cosmology,
Toronto, ON, M5G 1Z8 \\
$^{5}$Department of Astronomy \& Astrophysics, University of Toronto, 50 St George St, Toronto,
ON, M5S 3H4, Canada \\
$^{6}$Department of Physics \& Astronomy, University of British Columbia, 6224 Agricultural Rd., Vancouver, V6T 1Z1, Canada
}

\date{}

\pubyear{2017}

\begin{document}
\label{firstpage}
\pagerange{\pageref{firstpage}--\pageref{lastpage}}
\maketitle

\begin{abstract}
A new generation of wide-field radio interferometers designed for 21-cm surveys is being built as drift scan instruments allowing them to observe large fractions of the sky. With large numbers of antennas and frequency channels the enormous instantaneous data rates of these telescopes require novel, efficient, data management and analysis techniques. The $m$-mode formalism exploits the periodicity of such data with the sidereal day, combined with the assumption of statistical isotropy of the sky, to achieve large computational savings and render optimal analysis methods computationally tractable. We present an extension to that work that allows us to adopt a more realistic sky model and treat objects such as bright point sources. We develop a linear procedure for deconvolving maps, using a Wiener filter reconstruction technique, which simultaneously allows filtering of these unwanted components. We construct an algorithm, based on the Sherman-Morrison-Woodbury formula, to efficiently invert the data covariance matrix, as required for any optimal signal-to-noise weighting. The performance of our algorithm is demonstrated using simulations of a cylindrical transit telescope.
\end{abstract}




\section{Introduction}
\label{sec:intro}

Wide-field radio interferometers with large numbers of antennas and frequency channels are requiring increasingly complex and computationally intensive calibration, deconvolution, and imaging algorithms. In particular, 21-cm experiments such as BINGO \citep{bingoupdate},  CHIME \citep{chimepath1}, HIRAX \citep{hirax}, HERA \citep{hera}, and Tianlai \citep{tianlai} are drift scan telescopes -- pointing at a fixed location relative to the ground and allowing the sky to drift overhead -- providing large sky coverage in a simple and cost-effective manner (as they require few moving parts).

For synthesis imaging, typical map making procedures have been based on the CLEAN algorithm of \cite{clean}, which iteratively performs non-linear transformations on the input data to produce the output map. These can achieve computational efficiency by taking advantage of the locality of the inversion problem in map space, all while considering the direction-dependent and polarized nature of the primary beam, but have relied on techniques such as mosaicing to extend the field of view. Furthermore, a naive implementation of CLEAN is ill suited to the observation of diffuse structure, since its underlying assumption is that the sky is made up of point sources.

Alternatively, \cite{mmodes1, mmodes2} introduce a novel method for analyzing interferometric data of transit telescopes, the $m$-mode formalism. The method exploits the periodicity of such data with the sidereal day to achieve large computational savings, allowing the application of techniques developed for optimal analysis of Cosmic Microwave Background (CMB) data \citep{bondjaffeknox, tegmark, myerscbi} to be ported to full-sky 21-cm intensity mapping analysis \citep{liutegmark}. These techniques generally require diagonalization or inversion of the data covariance matrix as one of the most computationally intensive steps. The treatment of \cite{mmodes1, mmodes2} relies on the assumption of statistical isotropy of the sky to decompose this inversion into blocks. This assumption, however, is strongly broken in any realistic sky model, perhaps most visibly by the presence of a few bright radio point sources.

Here we adopt a more realistic sky model, relaxing the assumption of statistical isotropy, and show how we are able to treat such objects within the $m$-mode formalism. We develop a linear map making procedure, using a Wiener filter reconstruction technique, which simultaneously allows for deconvolution and point-source removal. Our algorithm, based on the Sherman-Morrison-Woodbury formula \citep{smf1, smf2, woodbury}, allows us to decompose the data covariance matrix into components and invert it with minimal perturbation to the large computational savings of the $m$-mode formalism.

In Section \ref{sec:mmodes} we give a brief digest of the $m$-mode formalism. In Section \ref{sec:model} we extend this treatment to a sky model that includes bright point sources and discuss optimal linear map making. In Section \ref{sec:algorithm} we present the Sherman-Morrison-Woodbury formula based algorithm for inverting the covariance matrix and filtering out the bright point source components. In Section \ref{sec:simulations}, we validate the technique on simulations of a mock observation and map making procedure. In Section \ref{sec:freq}, we analyze the frequency spectrum of these simulations. Finally, in Section \ref{sec:conclusions} we present our conclusions.

\section{\textit{m}-mode overview}
\label{sec:mmodes}

The $m$-mode formalism provides a convenient framework for discussing the measurement process of a transit interferometer on the full sky. We provide here a brief overview, highlighting the aspects and assumptions that allow one to isolate the relevant degrees of freedom and achieve large computational savings. For simplicity we consider only the case of an unpolarized sky. For the detailed discussion including the full polarized case we refer the reader to \cite{mmodes2}, although the results of Sections \ref{sec:model} and \ref{sec:algorithm} generalize straightforwardly. The visibilities $V_{ij}$ of a transit telescope are formed by the cross-correlation of the voltage signals received at antennas (or ``feeds'') $i$ and $j$, where $i,j=1,~\dots,~N_{\rm feeds}$,
\begin{equation}
V_{ij}(\phi) = \frac{1}{\sqrt{\Omega_{i}\Omega_{j}}} \int d^2\hat{n} ~ A_i(\nh; \phi) A^*_j(\nh; \phi) \, T(\nh) \, e^{2\pi i\nh \cdot \mathbf{u}_{ij}(\phi)},
\label{vis1}
\end{equation}
 where $\nh$ is the direction on the sky, $A_i$ is the antenna reception pattern (or ``beam''), $T$ is the sky brightness temperature in the Rayleigh-Jeans limit, and the quantity in the exponential is a geometric phase associated with the location of the source. For a detailed account of the measurement process in radio interferometry see \cite{radio1}. We normalize our visibilities by the geometric mean of the beam solid angle $\Omega_i =\int d^2\nh~|A_i({\nh})|^2$ so that for a sky with uniform brightness the auto-correlations $V_{ii}$ measure its temperature. Finally ${\bf u}_{ij} = (\mathbf{d}_{i} - \mathbf{d}_{j})/\lambda$, where $\mathbf{d}_{i}$ is the position vector of feed $i$ and $\lambda$ is the wavelength of observation. Although, for notational convenience, we have suppressed the frequency dependence in Eq. (\ref{vis1}), and furthermore in Sections \ref{sec:model} and \ref{sec:algorithm}, Section \ref{sec:freq} is devoted to its discussion.

For a transit telescope, the time dependence of the visibilities is periodic, following the rotation of the earth. This has allowed us, above, to replace the time dependence with the celestial azimuthal coordinate $\phi$. Since we would like to describe our measurement on the full sky, the $m$-mode formalism proceeds by decomposing in spherical harmonics
\begin{align}
 B_{ij}(\nh; \phi) &\equiv  \frac{1}{\sqrt{\Omega_{i}\Omega_{j}}} \, A_i(\nh; \phi) A^*_j(\nh; \phi) \, e^{2\pi i\nh \cdot \mathbf{u}_{ij}(\phi)}
 \nonumber \\
 &= \sum_{l,m} B_{ij,lm}(\phi) \, Y_{lm}^* (\nh),
 \label{btm}
 \\
  T(\nh) &= \sum_{l,m} a_{lm} \, Y_{lm}(\nh),
 \end{align}
 where in Eq. (\ref{btm}) we have defined the beam transfer functions $B_{ij}(\nh; \phi)$ and their spherical harmonic coefficients, the beam transfer matrices $B_{ij,lm}(\phi)$. Due to the periodicity of our signal with respect to $\phi$ we may take its Fourier transform to obtain the $m$-mode visibilities
\begin{equation}
V_{ij,m} = \int d\phi ~ V_{ij}(\phi) \, e^{-im\phi},
\end{equation}
and noticing that the $\phi$ dependence of the transfer function simply rotates it about the polar axis $B_{ij,lm}(\phi) = B_{ij,lm}(\phi = 0) e^{im\phi}$, the sum over $m$ collapses giving
\begin{equation}
V_{ij,m} = \sum_{l} B_{ij,lm} \, a_{lm}.
\label{mvis}
\end{equation}
Equation (\ref{mvis}) says that the measurement process of a transit telescope does not mix $m$-modes. We see that we have isolated the relevant degrees of freedom since the response of an interferometer is compact in $m$-space, limited by the East-West extent of the array
\begin{equation}
m_{\rm max} = 2\pi \frac{d_{\rm EW}}{\lambda}.
\end{equation}
We may also write Eq. (\ref{mvis}) in explicit matrix notation
\begin{align}
(\mathbf{v}_m)_{(ij)} = \left( \mathbf{B}_m \right)_{(ij)(l)}(\mathbf{a}_m)_{(l)} + (\mathbf{n}_m)_{(ij)},
\label{matrix}
\end{align}
where, for example for matrices, two sets of parentheses are used to separate indices that contribute to either rows or columns. So, for the beam transfer matrices, we have
\begin{align}
\left( \mathbf{B}_m \right)_{(ij)(l)}  =  B_{ij,lm}.
\end{align}
In the following, when employing explicit matrix notation we will use the convention that repeated indices are summed over. In Eq. \eqref{matrix}, we have also included a contribution from instrumental noise, which we assume to be uncorrelated between antennas and frequencies. We see that the $m$-independence of Eq. (\ref{mvis}) is realized as matrices that are block diagonal in $m$.

Furthermore, if we assume that the sky is a statistically isotropic random field then its covariance matrix
\begin{equation}
(\mathbf{C})_{(lm)(l'm')} \equiv \langle a_{lm} a_{l'm'}^* \rangle = \delta_{ll'} \delta_{mm'} C_l,
\end{equation}
is also block diagonal in $m$. Since each $m$-block may be treated independently during the analysis process, this renders the tasks of optimal linear map making, foreground removal, and quadratic power spectrum estimation computationally tractable. It provides savings of order $m_{\rm max}^2$ in diagonalizing or inverting the covariance matrix, and allows the computation to be distributed across many parallel processes.

\section{Sky model and Wiener filter technique}
\label{sec:model}

\begin{figure*}
\centering
\includegraphics[width=\textwidth]{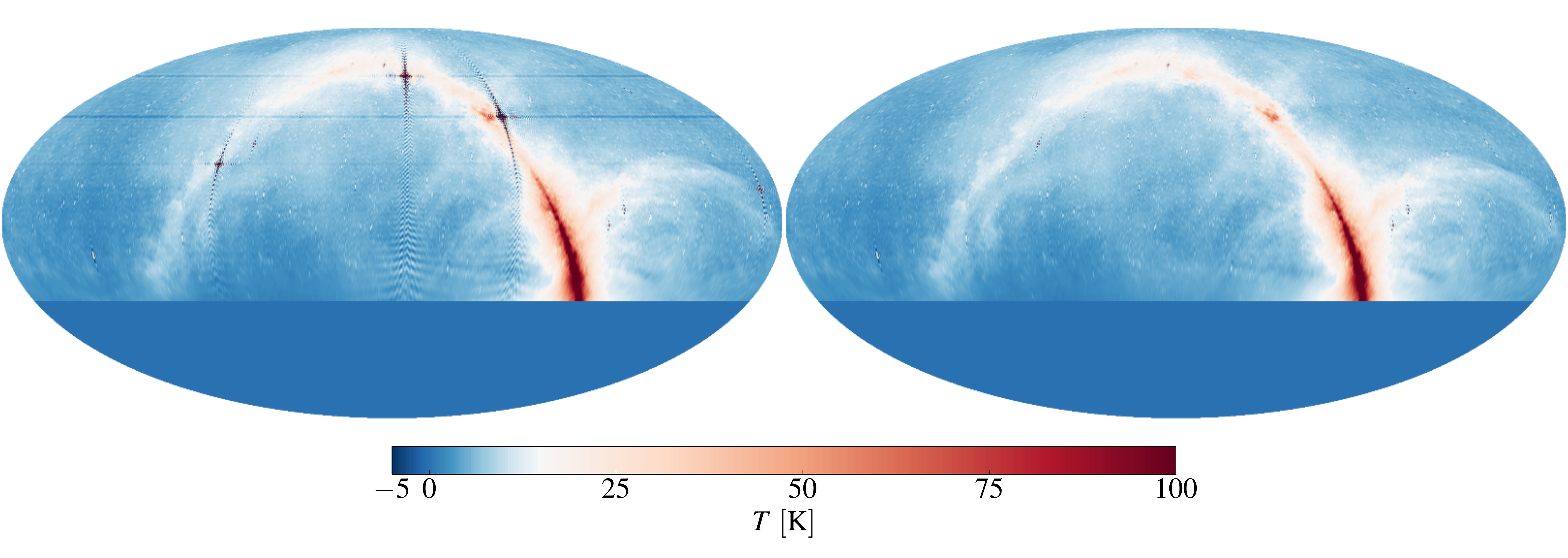}
\caption{A simulated CHIME Pathfinder reconstruction of the smooth component at $624$ MHz using the Wiener filter formalism described in Section \ref{sec:model}. The bottom section of the map is masked since it is below the horizon of the simulated telescope. (left) The deconvolved map with no point-source removal, equivalent to setting the amplitude of the point source covariance in Eq (\ref{filter}) to zero. (right) The deconvolved map where the four brightest point sources have been removed. The colour bar is a linear scale between $-5$ and $100~{\rm ^{\circ}K}$.}
\label{standard}
\end{figure*}

The true radio sky is by no means statistically isotropic. At low frequencies and multipoles the diffuse sky is dominated by synchrotron emission from the Milky Way, which varies as a function Galactic latitude. Furthermore, a few bright point sources dominate the dynamic range and therefore can cause significant non-local artefacts in a naive attempt to deconvolve a map. Therefore, for this illustration, we consider a model with two components
\begin{equation}
\mathbf{a} = \mathbf{a}_{\rm sm} + \mathbf{a}_{\rm kps}.
\end{equation}
$\mathbf{a}_{\rm sm}$ is a smooth component, meaning its covariance matrix $\mathbf{S}_{\rm sm} = \langle \mathbf{a}_{\rm sm}\mathbf{a}_{\rm sm}^\dagger \rangle$ is characterized by a negative power-law index in $l$.
 $\mathbf{a}_{\rm kps}$ is the contribution from a collection of known point sources. We would like to construct an estimate of the smooth component $\hat{\mathbf{a}}_{\rm sm}$ and so filter out the point sources with a linear operation on the data \citep{wiener}
\begin{equation}
\hat{\mathbf{a}}_{\rm sm} = \mathbf{F}\mathbf{v}.
\label{estimate}
\end{equation}
We look for a filter $\mathbf{F}$ that minimizes the variance of the residuals between the estimate and the signal
\begin{equation}
\frac{\delta}{\delta \mathbf{F}} \left\langle (\mathbf{a}_{\rm sm} - \mathbf{F}\mathbf{v})^{\dagger} (\mathbf{a}_{\rm sm} - \mathbf{F}\mathbf{v}) \right\rangle = 0,
\label{minimize}
\end{equation}
where angled brackets denote an ensemble average of realizations of the sky $\mathbf{a}_{\rm sm}$ and $\mathbf{a}_{\rm kps}$, and of the instrumental noise $\mathbf{n}$. Assuming that the cross-correlations between the two signal components is zero, as are their cross-correlations with the noise, the solution to equation (\ref{minimize}) is
\begin{align}
\mathbf{F} &= \langle \mathbf{a}_{\rm sm}\mathbf{v}^\dagger \rangle \langle \mathbf{v}\mathbf{v}^{\dagger} \rangle^{-1} \\
&= \mathbf{S}_{\rm sm}\mathbf{B}^{\dagger}\left( \mathbf{N} + \mathbf{B}(\mathbf{S}_{\rm sm} + \mathbf{S}_{\rm kps})\mathbf{B}^{\dagger}\right)^{-1}, \label{filter}
\end{align}
where $\mathbf{N}$ is the noise covariance matrix. Combining Eqs. (\ref{filter}) and (\ref{estimate}) we obtain the optimal Wiener filtered reconstruction of the smooth component, which requires inversion of the covariance matrix of the visibilities. However, contrary to $\mathbf{N}$ and $\mathbf{S}_{\rm sm}$, $\mathbf{S}_{\rm kps}$ cannot be assumed diagonal in $m$. In general, it is given by an ensemble average over the outer product of the spherical harmonic coefficients of the measured point sources,
\begin{equation}
\mathbf{(S)}^{\rm kps}_{(lm)(l'm')} = \left\langle \mathbf{a}^{\rm kps}_{(lm)}\mathbf{a}^{{\rm kps}*}_{(l'm')}\right\rangle.
\label{kpscov}
\end{equation}
Although, if we wish to consider the collection as a single component, we must assume that their relative brightnesses are known exactly, so that only their overall amplitude relative to the smooth component must be fit. In this case, the covariance matrix of the known point sources becomes rank-1 (See Section \ref{sec:multicomponent} for the generalization to the multi-component case). As well, in our model, we must assume the positions of the point sources are known exactly, or are determined to very high accuracy in radio surveys so that their uncertainties are negligible. Although we do not consider the effect of position uncertainties in this work, it has been studied recently in the context of foreground removal for 21-cm intensity mapping \citep{barry2016, ewall-wice2016}. 

Note that this method can be generalized trivially to any single component, such as extended sources whose spatial distribution is well-known (or can be decomposed into components, referring again to Section \ref{sec:multicomponent}). In this context, perfect point sources with accurate positioning can be considered the ideal case. 

Of course, the contribution of the known point sources does not have zero mean. Still, we do not subtract them explicitly. We simply set the amplitude of their covariance to be large relative to the smooth component, allowing us to project out all modes associated with them, by down weighting after projecting into visibilities (Eq. \eqref{filter}). Similar methods have been used successfully, for example, in analysis of CMB interferometric data \citep{myerscbi}.

\section{Sherman-Morrison-Woodbury formula based algorithm}
\label{sec:algorithm}

\begin{figure*}
\centering
\includegraphics[width=\textwidth]{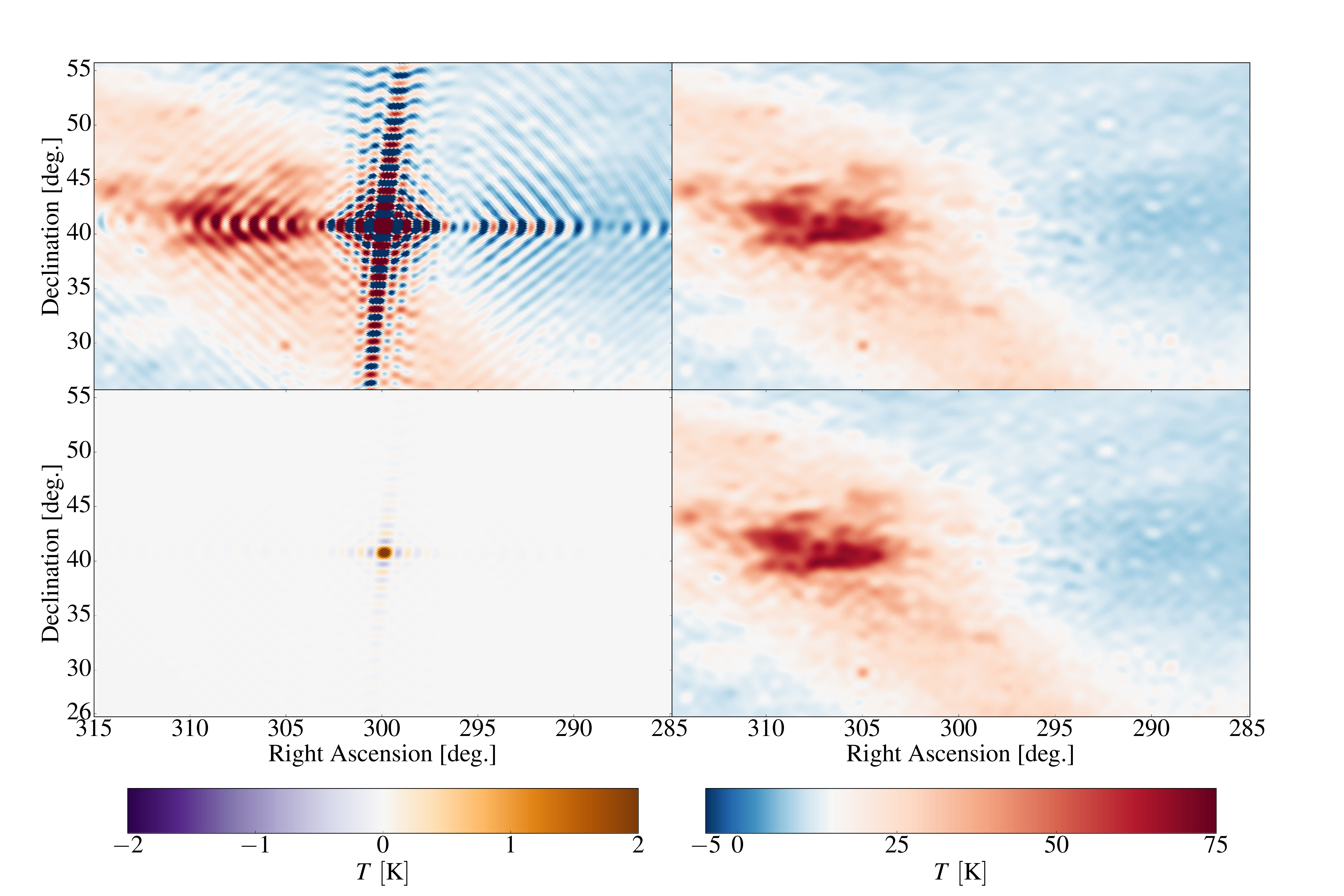}
\caption{The Cygnus A region of the map shown in Figure \ref{standard} at 624 MHz, showing the comparison between a standard Wiener filter (top left), one where Cygnus A has been projected out (top right), a standard Wiener filter on a reference simulated map that never had Cygnus A to begin with (bottom right), and the residuals between the deconvolution and reference (bottom left). The bottom right temperature scale, a linear scale between $-5$ and $75~{\rm ^{\circ}K}$, applies to all panels except the residuals (bottom left), whose colour scale is directly below it. The residuals are seen to be point-source-like, local to the region near the point source, smaller than the simulated diffuse signal component, and $\sim 3$ orders of magnitude smaller than the artefacts in the top left panel. Simulated visibilities are produced using an idealized model of the CHIME Pathfinder telescope.}
\label{cygzoom}
\end{figure*}

\subsection{Two component case}
\label{sec:singlecomponent}

Unfortunately, to make use of the method of the previous section, the presence of bright point sources means we must invert a dense $(l_{\rm max}m_{\rm max} \times l_{\rm max}m_{\rm max})$-dimensional matrix. Whereas previously, with the assumption of statistical isotropy, the inversion could be broken into $m$-blocks allowing the computation to be spread across many parallel processes and the matrix held distributed in memory, now information from all $m$-blocks must be shared. However, to be specific, the covariance of the known point sources (\ref{kpscov}) is a rank-1 matrix. Taking advantage of this fact, the inversion can still be computed within the framework of $m$-modes, and with minimal communication, with the aid of the Sherman-Morrison formula \citep{smf1, smf2}
\begin{equation}
(\mathbf{A} + \mathbf{uu}^{\dagger})^{-1} = \mathbf{A}^{-1} - \frac{\mathbf{A}^{-1}\mathbf{u}\mathbf{u}^{\dagger}\mathbf{A}^{-1}}{1+\mathbf{u}^{\dagger}\mathbf{A}^{-1}\mathbf{u}}, \label{smf}
\end{equation}
where $\mathbf{A}$ is an $(M \times M)$--dimensional matrix and $\mathbf{u}$ is a vector of length $M$.
Equation (\ref{filter}) can be cast in the form of Eq. (\ref{smf}) if we identify
\begin{align}
\mathbf{A}  = \left( \mathbf{N} + \mathbf{B}\mathbf{S}_{\rm sm}\mathbf{B}^{\dagger}\right),
\end{align}
and
\begin{align}
\mathbf{u} = \mathbf{B}\mathbf{a}_{\rm kps}.
\end{align}
The first piece is block diagonal in $m$, so $\mathbf{A}^{-1}$ can be computed on an $m$-by-$m$ basis. However all of $\mathbf{A}^{-1}$ is required to multiply $\mathbf{u}$ and form the outer product on the right hand side of Eq. (\ref{smf}). Notice though that we would actually like to compute
\begin{align}
\hat{\mathbf{a}}_{\rm sm} &= \mathbf{S}_{\rm sm}\mathbf{B}^{\dagger}\left( \mathbf{N} + \mathbf{B}(\mathbf{S}_{\rm sm} + \mathbf{S}_{\rm kps})\mathbf{B}^{\dagger}\right)^{-1}\mathbf{v}
\nonumber \\
&= (\mathbf{S}_{\rm sm}\mathbf{B}^{\dagger})_{(lm)(ijm')}
\Bigg[ (\mathbf{A}^{-1}\mathbf{v})_{(ijm')}
\nonumber \\
&- \underbrace{\frac{1}{1+\mathbf{u}^{\dagger}\mathbf{A}^{-1}\mathbf{u}}}_{\text{\RNum{1}}}
 \underbrace{\vphantom{\frac{1}{1+\mathbf{u}^{\dagger}\mathbf{A}^{-1}\mathbf{u}}}(\mathbf{A}^{-1}\mathbf{u})_{(ijm')}}_{\text{\RNum{3}}}
  \underbrace{\vphantom{\frac{1}{1+\mathbf{u}^{\dagger}\mathbf{A}^{-1}\mathbf{u}}}(\mathbf{A}^{-1}\mathbf{u})^*_{(i'j'm'')} \mathbf{v}_{(i'j'm'')}}_{\text{\RNum{2}}}
\Bigg].
\label{expand}
\end{align}
There are therefore three pieces in the rank-1 correction to the final map that need to computed, consisting of only matrix or vector multiplications: the normalisation in the denominator (\RNum{1}), the scalar in the numerator made up of the inner product of the known point source map and the visibilities (\RNum{2}), and the vector that is some weighted version of the known point source map (\RNum{3}). Pieces (\RNum{1}) and (\RNum{2}) are sums over $m$, each of which can be broken up, performed in $m$-blocks, and then summed across (the only step requiring communication between $m$s). The vector piece (the correction in map space once (\RNum{3}) is multiplied by $\mathbf{S}_{\rm sm}\mathbf{B}^{\dagger}$) can easily be saved in memory as each piece is computed over $m$, then shared in the same step as the normalisation.

\subsection{Multi-component case}
\label{sec:multicomponent}

Consider the case where we would like to treat many components as statistically independent. For example, we might be interested in fitting the amplitudes of many point sources as observed in the data and then removing them separately. Assuming there is no cross-correlation between the various components, we can generalize the results of the previous section by simply performing the Sherman-Morrison formula iteratively on each rank-1 component of the covariance matrix. This extension allows for the inversion of an arbitrary number of components or rank of covariance matrix, with no increase in the memory footprint, at the expense of one communication step per component.

Alternatively, we may rewrite our measurement equation (\ref{matrix}) as
\begin{equation}
\mathbf{v} =\mathbf{B}\mathbf{a}_{\rm sm} + \mathbf{B}\mathbf{P}\mathbf{s}_{\rm kps}  + \mathbf{n},
\end{equation}
where $\mathbf{s}_{\rm kps}$ is a vector of size $N_{\rm kps}$, the number of known point sources, containing their fluxes and $\mathbf{P}$ is a projection matrix which maps these fluxes to their locations on the sky, sums the components, and then performs a spherical harmonic transform. The task of inverting the covariance matrix of the collection of point sources is nicely decomposed using the Woodbury matrix identity (\citealp{woodbury}; of which the Shermann-Morrison formula is a special case)
\begin{equation}
(\mathbf{A}+\mathbf{U}\mathbf{C}\mathbf{V})^{-1}=\mathbf{A}^{-1}-\mathbf{A}^{-1}\mathbf{U}(\mathbf{C}^{-1}+\mathbf{V}\mathbf{A}^{-1}\mathbf{U})^{-1}\mathbf{V}\mathbf{A}^{-1}.
\label{woodbury}
\end{equation}
Similarly to the previous section, we may identify $\mathbf{U} = \mathbf{B}\mathbf{P}$, $\mathbf{V} = \mathbf{U}^\dagger$, and $\mathbf{C} = \langle \mathbf{s}_{\rm kps}\mathbf{s}_{\rm kps}^{\dagger} \rangle$. Our estimate of the smooth component becomes
\begin{align}
\hat{\mathbf{a}}_{\rm sm} &=  \mathbf{S}_{\rm sm}\mathbf{B}^{\dagger}\Big( \mathbf{A}^{-1}\mathbf{v}
\nonumber \\
 &-  \underbrace{\vphantom{\big]^{-1}_{(n)(n')}}(\mathbf{A}^{-1}\mathbf{U})_{(ijm)(n)}}_{\text{\RNum{3}}}
 \underbrace{\big[\mathbf{C}^{-1}
+\mathbf{U}^\dagger\mathbf{A}^{-1}\mathbf{U}\big]^{-1}_{(n)(n')}}_{\text{\RNum{1}}}
\nonumber \\
&\times
\underbrace{\vphantom{\big]^{-1}_{(n)(n')}}(\mathbf{U}^{\dagger}\mathbf{A}^{-1})_{(n')(i'j'm')}\mathbf{v}_{(i'j'm')}}_{\text{\RNum{2}}} \Big),
\label{expand2}
\end{align}
where $n=1,\dots,N_{\rm kps}$. We see from (\RNum{3}) in Eq. (\ref{expand2}) that, while in Section \ref{sec:singlecomponent} we needed to hold only a single vector in memory during the computation, we must now hold $N_{\rm kps}$ vectors. However, the entire inversion can be completed with only two steps of communication. The multiplication of $\mathbf{A}^{-1}$ with $\mathbf{U}=\mathbf{B}\mathbf{P}$ or its conjugate mixes all $m$-modes. Therefore, the sum inside the square brackets (\RNum{1}) must first be computed block-wise and the result shared. The resulting matrix is only $(N_{\rm kps} \times N_{\rm kps})$ and so can be inverted easily on each processor after the first communication. Finally, the piece of the multiplication of that matrix with the visibilities ((\RNum{1}) with (\RNum{2})) needs to be computed and shared to produce the final correction. We see that this algorithm provides computational savings over the brute force inversion as long as $N_{\rm kps} < m_{\rm max}(m_{\rm max} - 1)/2$. In practice, we expect $N_{\rm kps}$ to be on the order of a few tens, which drastically improves the performance, since $m_{\rm max}$ is generally at least few hundred.
\section{Simulations}
\label{sec:simulations}

To validate our technique we perform a mock observation and map-making procedure on simulated data. Using the tools provided in the \texttt{driftscan}, \texttt{caput}, and \texttt{cora} packages\footnote{\url{https://github.com/radiocosmology}}, we are able to generate a simulated radio sky and then produce visibilities using an idealized model of the CHIME Pathfinder telescope \citep{chimepath1}. The CHIME Pathfinder, located at latitude 49.3 degrees, consists of two 37-m $\times$ 20-m long paraboloidal cylinders oriented north-south. Each is instrumented with 64 dual-polarisation dipole antennas spaced by 0.3 m. The simulated sky model, described by \cite{mmodes2}, contains both diffuse Galactic emission and point sources. The diffuse emission consists of an extrapolated Haslam map \citep{haslam1981, haslam1982} with angular and spectral fluctuations added to unconstrained scales and frequencies. The point source signal is constructed from a catalog of known sources (brighter than 10~Jy at 151~MHz), a synthetic catalog of fainter sources (brighter than 0.1~Jy), and a random background of unresolved sources below that. It also contains a model of polarisation for the diffuse and point source components, whose details we will not include here. After the mock observation, we proceed to deconvolved maps following our point-source removal algorithm, as a custom implementation of the standard Wiener filter map making algorithm provided in \texttt{draco}. For these simulations, the set of beam transfer matrices that are used to produce the visibilities are the same used during the map making. In practice, when applied to the data, our ability to project out point sources will be limited by the accuracy of the beam transfer matrices. Therefore this technique both relies on, and allows to probe one's understanding of, the model for the point spread function of the telescope.

\begin{figure}
\centering
\includegraphics[width=0.5\textwidth]{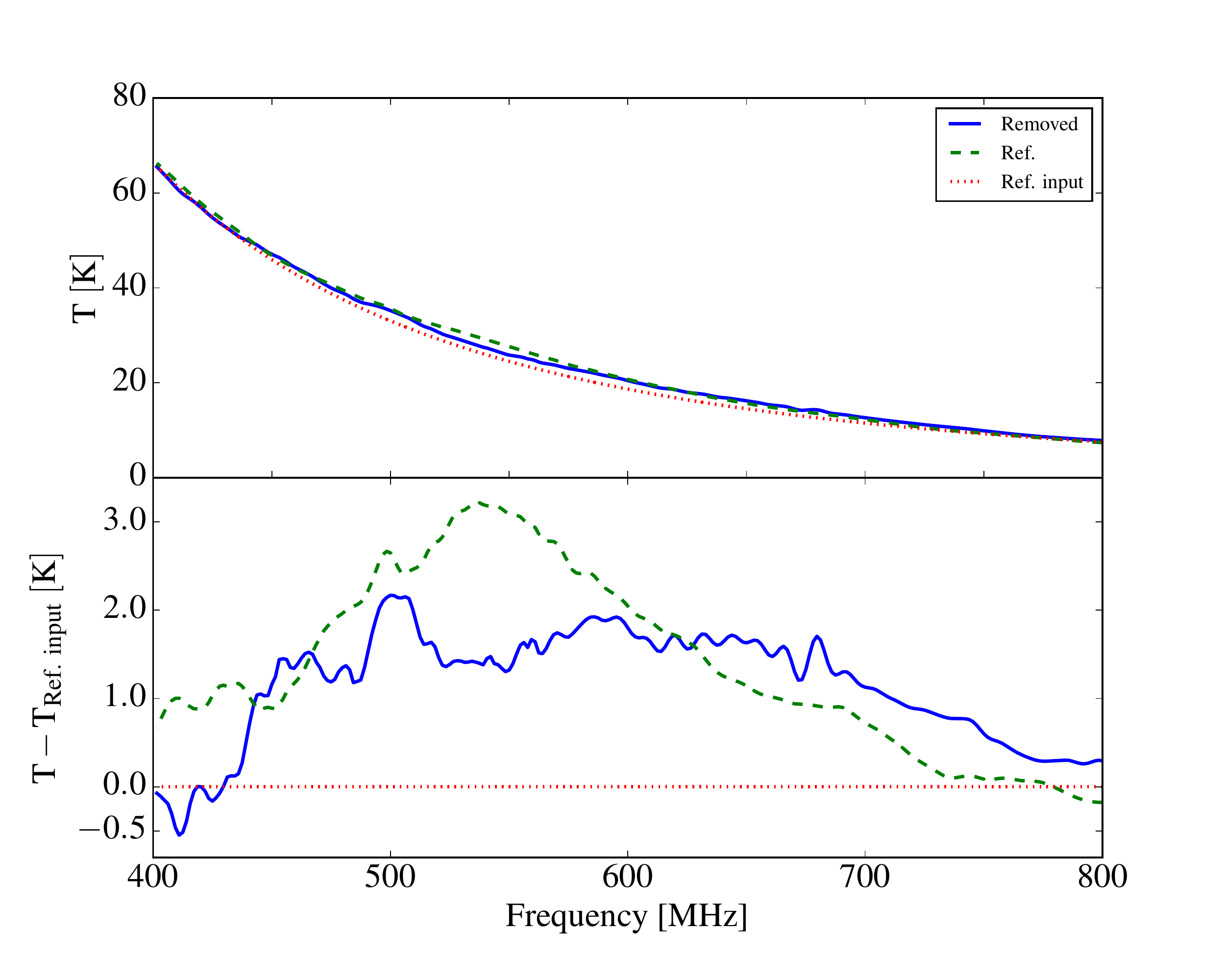}
\caption{Intensity spectra between 400 and 800 MHz at the location of Cygnus A, computed by performing the corresponding Wiener filter independently at each frequency. The blue solid line shows the output of our method, while the reference simulation (green dashed) performed a standard Wiener filter on an input map (red dotted) that already had Cygnus A removed exactly.}
\label{residuals}
\end{figure}

For illustration we simulate the removal of a single component consisting of the four brightest radio point sources (Cygnus A, Cassiopeia A, Taurus A, and Virgo A). We inform the model by using a point source map as input that contains only four non-zero pixels at the exact locations of the sources in the simulated sky used to generate the visibilities. Figures \ref{standard} and \ref{cygzoom} show comparisons between a Wiener filter map with these four bright sources removed, and one where no point-source removal has been applied. The latter is equivalent to setting the amplitude of the known point source covariance in the model of Section \ref{sec:model} to zero. Figure \ref{standard} shows the full sky deconvolved map with and without removal of the four brightest point sources. 

In order to assess the performance of our method, we compare our point-source-removed maps to a reference simulation, that is otherwise identical except it never had the point source in it to begin with and no point-source removal is applied in the deconvolution. We may then study the residuals by subtracting the reference from the point-source-removed map. Figure \ref{cygzoom} shows a zoom into the Cynus A region of this comparison, for the map shown in Figure \ref{standard} (for a single frequency). The bottom left panel shows that the central region of the point source is positive, indicating that some ammount of it has not been fully removed. However this is not true for all frequencies. In Figure \ref{residuals} we show the spectrum of these residuals at the central location of the point source between 400 and 800 MHz, when the deconvolution is computed independently at each frequency and only Cygnus A is removed. We employ an agnostic prior of $\mathbf{S}_{\rm sm} \propto \delta_{\nu\nu'}\nu^{-2}$ for the spectral tilt of the smooth component, although we find the results do not qualitatively depend on this choice. We discuss the results of the full spectrum calculations and their implications in the following section.

\section{Frequency structure of the covariance}
\label{sec:freq}

In Section \ref{sec:simulations}, we discussed the method used to analyze the performance of our method: A comparison to a reference simulation that never had the point source to begin with. Figure \ref{residuals} shows the results of the calculation, comparing the spectrum (between 400 and 800 MHz) at the location of the point source in the removed simulation to the reference, and to the reference sky before observation (which consists of diffuse Galactic emission). The spectra are produced by performing the deconvolution on each frequency independently. When the deconvolution is performed this way, significant frequency structure is introduced into the estimate of the signal, for both the removed and reference maps. This is a realization of an effect known as mode mixing (although in a somewhat novel context, since this is mode mixing in the optimal estimate): The telescope beam, whose spatial extent varies as a function of frequency, beats against the larger-scale structure of the surrounding diffuse emission. While for the simulations presented here, the residuals are well below the level of the estimated signal (a $\sim 3$ orders of magnitude reduction of the original point source flux), such structure has been seen to be problematic in the context of 21-cm intensity mapping, where the foreground filter depends on the assumption of smooth-spectrum foregrounds \citep{barry2016}.

Clearly the assumption of independent frequencies is inadequate. We would like to dampen the high-frequency oscillations in the estimate by informing our model that neighboring frequencies are actually correlated, that is by introducing a covariance matrix which is non-diagonal in frequency. One option is to assume that the point source should be perfectly correlated in frequency, in which case the results of Section \ref{sec:algorithm} can be applied to the full set of visibilities, but this may be too strict. Unfortunately, the deconvolution problem for a general frequency covariance matrix poses a computational predicament. To understand why, consider the case where the covariance matrices of the signal and point source factorize as tensor products, $\mathbf{S}_{\rm sm} = \mathbf{A} \otimes\mathbf{G}$ and $\mathbf{S}_{\rm kps} = \mathbf{u}\mathbf{u}^\dagger \otimes\mathbf{H}$. Matrices on the left of the tensor product then carry the spatial dependence, while those on the right are the frequency parts. We could attempt the inversion using the equation
\begin{align}
\left(\mathbf{A} \otimes \mathbf{I} + \mathbf{u}\mathbf{u}^\dagger \otimes \mathbf{H} \right)^{-1} = ~& \mathbf{A}^{-1} \otimes \mathbf{I} \label{PRODUCT}
\\ &- \mathbf{A}^{-1}\mathbf{u}\mathbf{u}^\dagger\mathbf{A}^{-1} \otimes \left(\mathbf{I} + \mathbf{u}^\dagger\mathbf{A}^{-1}\mathbf{u} \mathbf{H}\right)^{-1} \mathbf{H} \nonumber
\end{align}
(for a proof see Appendix \ref{sec:proof}), but we immediately encounter the standard issue that the signal and noise covariance matrices are diagonal in different bases.\footnote{Eq. \eqref{PRODUCT} solves the case of $\mathbf{G}=\mathbf{I}$, where $\mathbf{I}$ is the identity, but the case for general $\mathbf{G}$ can easily be solved by simply factoring it, assuming it is invertible, which simply modifies the definition of $\mathbf{H}$.} Even in the confusion limit, we can only expect this factorizability for on-sky quantities, which must then be tranformed into the data basis by multplication with beam transfer matrices. Considering the combined frequency and spatial coordinate matrices in block matrix form, this multiplication applies a smoothly varying function across the diagonal of the tensor product structure, forcing us to consider the full shape of the covariance matrix.

A simple solution is to enforce a smooth spectrum of the removed point source by performing a low-order polynomial fit to the point source estimated with a diagonal-in-frequency covariance matrix before removal. This method has been shown to be effective for the purpose of 21-cm intensity mapping by \cite{barry2016}. Alternatlively, due to the highly structured form of the covariance matrix, a conjugate gradient scheme for performing the full inversion -- with the results of this work as the initial guess -- could be feasible, although we leave this for later work.

\section{Conclusion}
\label{sec:conclusions}
We have shown that the $m$-mode formalism of \cite{mmodes1, mmodes2} is indeed convenient for analysis of full-sky data from transit interferometers, even when the assumption of statistical isotropy of the sky is relaxed. In Section \ref{sec:model} we adopted a realistic sky model that includes components, namely (although not limited to) bright point sources, which display the relationship between statistical anisotropy and a covariance matrix that is non-diagonal in $m$. Furthermore, in Section \ref{sec:simulations}, we demonstrated with simulations how such components cause heavily non-local ringing in a standard attempt to optimally estimate a map. We then showed how one can use a linear, optimal Wiener filter reconstruction technique to project out the components in the same step as the deconvolution and map making. For the task of inverting the covariance matrix, we developed an algorithm based on the Sherman-Morrison-Woodbury formula (Section \ref{sec:algorithm}), which adds minimal computational cost to a method that assumes a block diagonal in $m$ structure. Indeed, as mentioned at the end of Section \ref{sec:model}, this algorithm can be used to efficiently either estimate or project-out any low-rank component for which the diagonalization transformation is known. In Section \ref{sec:simulations}, we described the simulation technique used to validate our method and computed the spectrum of residuals, which were shown to be small compared to the estimated signal. In Section \ref{sec:freq}, we discussed the computational difficulty of considering the full frequency structure of the data covariance matrix, but suggested several viable solutions. This method will be useful, for example, for efforts to map the synchrotron emission of the Milky Way or the cosmic 21-cm intensity, on the full-sky.

\section*{Acknowledgments}

We wish to thank T. Landecker for having (so quickly) reviewed and provided useful comments on the original manuscript. We also wish to thank N. T. Denman, L. Newburgh, and K. Vanderlinde for useful comments throughout. Finally, we would like to thank the anonymous reviewer for his/her comments which lead to the development of Section \ref{sec:freq}.  Computations were performed on the General Purpose Cluster supercomputer at the SciNet HPC Consortium. \citep{scinet} SciNet is funded by: the Canada Foundation for Innovation under the auspices of Compute Canada; the Government of Ontario; Ontario Research Fund - Research Excellence; and the University of Toronto.

\appendix
\section{Proof of equation (\ref{PRODUCT})} \label{sec:proof}

We can show Eq. \eqref{PRODUCT} by direct multiplication,
\begin{align}
&\left(\mathbf{A} \otimes \mathbf{I} + \mathbf{u}\mathbf{u}^\dagger \otimes \mathbf{H} \right) 
 \left( \mathbf{A}^{-1} \otimes \mathbf{I} - \mathbf{A}^{-1}\mathbf{u}\mathbf{u}^\dagger\mathbf{A}^{-1} \otimes \left(\mathbf{I} + \mathbf{u}^\dagger\mathbf{A}^{-1}\mathbf{u} \mathbf{H}\right)^{-1} \mathbf{H} \right)
 \nonumber \\
 &= \mathbf{I} \otimes \mathbf{I} + \mathbf{u}\mathbf{u}^\dagger\mathbf{A}^{-1} \otimes \mathbf{H} - \mathbf{u}\mathbf{u}^\dagger\mathbf{A}^{-1}\otimes \left(\mathbf{I} + \mathbf{u}^\dagger\mathbf{A}^{-1}\mathbf{u} \mathbf{H}\right)^{-1} \mathbf{H}
\nonumber \\
&~- \mathbf{u}\mathbf{u}^\dagger\mathbf{A}^{-1}\mathbf{u}\mathbf{u}^\dagger\mathbf{A}^{-1}\otimes \mathbf{H} \left(\mathbf{I} + \mathbf{u}^\dagger\mathbf{A}^{-1}\mathbf{u} \mathbf{H}\right)^{-1} \mathbf{H}. \label{A1}
\end{align}
The product on the last line of Eq. \eqref{A1} has formed a scalar, which can be brought through the tensor product,
\begin{align}
&= \mathbf{I} \otimes \mathbf{I} + \mathbf{u}\mathbf{u}^\dagger\mathbf{A}^{-1} \otimes \mathbf{H} - \mathbf{u}\mathbf{u}^\dagger\mathbf{A}^{-1}\otimes \left(\mathbf{I} + \mathbf{u}^\dagger\mathbf{A}^{-1}\mathbf{u} \mathbf{H}\right)^{-1} \mathbf{H}
\nonumber \\
&~- \mathbf{u}\mathbf{u}^\dagger\mathbf{A}^{-1}\otimes \mathbf{u}^\dagger\mathbf{A}^{-1}\mathbf{u}\mathbf{H} \left(\mathbf{I} + \mathbf{u}^\dagger\mathbf{A}^{-1}\mathbf{u}\mathbf{H}\right)^{-1} \mathbf{H}. 
\end{align}
The last two terms can then be factored,
\begin{align}
=& \mathbf{I} \otimes \mathbf{I} + \mathbf{u}\mathbf{u}^\dagger\mathbf{A}^{-1} \otimes \mathbf{H} 
\\ \nonumber &- \left( \mathbf{I} \otimes \mathbf{I} + \mathbf{I} \otimes \mathbf{u}^\dagger\mathbf{A}^{-1}\mathbf{u} \mathbf{H}\right)
 \left( \mathbf{u}\mathbf{u}^\dagger\mathbf{A}^{-1} \otimes \left(\mathbf{I} + \mathbf{u}^\dagger\mathbf{A}^{-1}\mathbf{u}\mathbf{H}\right)^{-1} \mathbf{H}\right)
\\ =& \mathbf{I} \otimes \mathbf{I} + \mathbf{u}\mathbf{u}^\dagger\mathbf{A}^{-1} \otimes \mathbf{H} 
\\ \nonumber &- \left( \mathbf{I} \otimes \mathbf{I} + \mathbf{u}^\dagger\mathbf{A}^{-1}\mathbf{u} \mathbf{H} \right)
 \left( \mathbf{u}\mathbf{u}^\dagger\mathbf{A}^{-1} \otimes \left(\mathbf{I} + \mathbf{u}^\dagger\mathbf{A}^{-1}\mathbf{u}\mathbf{H}\right)^{-1} \mathbf{H}\right).
 \end{align}
 Multiplying through we obtain
 \begin{align}
 =& \mathbf{I} \otimes \mathbf{I} + \mathbf{u}\mathbf{u}^\dagger\mathbf{A}^{-1} \otimes \mathbf{H} \nonumber 
\\ &- \mathbf{u}\mathbf{u}^\dagger\mathbf{A}^{-1} \otimes \mathbf{H}.
\end{align}
The result can similarly be shown for multiplication on the left.




\bibliographystyle{mnras}
\bibliography{PSesandMmodes}


\bsp	
\label{lastpage}
\end{document}